\documentclass[fleqn,usenatbib]{mnras}

\usepackage{newtxtext,newtxmath}

\usepackage[T1]{fontenc}

\DeclareRobustCommand{\VAN}[3]{#2}
\let\VANthebibliography\thebibliography
\def\thebibliography{\DeclareRobustCommand{\VAN}[3]{##3}\VANthebibliography}

\usepackage{graphicx}	
\usepackage{amsmath}	
\usepackage{amssymb}	

\def\mnras{MNRAS}
\def\apj{ApJ}
\def\aj{AJ}

\def\aap{A\&A}
\def\apjl{ApJL}
\def\apjs{ApJS}

\def\araa{ARA\&A}
\def\nat{Nature}

\def\pasp{Pub. Ast. Soc. Pacific}

\def\bhm{M_{\bullet}}

\def\calL{{\cal J}}
\def\calM{{\cal M}}
\def\calR{{\dot{\cal{R}}}}

\def\calT{{\cal T}_{\rm tid}}
\def\cblue{}

\def\cs{c_{\rm s}}
\def\deltaR{\delta R}

\def\dotm{\dot{m}}

\def\ergs{\rm ergs\,s^{-1}}
\def\etal{{\it et al.}}

\def\feii{Fe\,{\sc ii}}

\def\kms{\rm km\,s^{-1}}

\def\NGC{NGC\,1068}

\def\oiii{[O\,{\sc iii}]}

\def\Rs{R_{\rm Sch}}

\def\RKHI{R_{\rm KHI}}

\def\sunm{M_{\odot}}

\def\tK{t_{\rm K}}
\def\VK{v_{\rm K}}
\def\VR{v_{\rm R}}

\begin{document}

\voffset=-0.35in

\title[{Close Binary of Supermassive Black Holes in \NGC}]
{Dynamical evidence from the sub-parsec counter-rotating  disc for a close binary of supermassive 
black holes in NGC\,1068}

\author[Wang et al.]
{Jian-Min Wang$^{1,2,3}$\thanks{E-mail: wangjm@ihep.ac.cn}, 
Yu-Yang Songsheng$^{1,2}$,
Yan-Rong Li$^{1}$,
Pu Du$^{1}$ and
Zhe Yu$^{1,2}$ 
\\ \\
$^{1}$Key Laboratory for Particle Astrophysics, Institute of High Energy Physics,
Chinese Academy of Sciences, 19B Yuquan Road, Beijing 100049, China\\
$^{2}$School of Astronomy and Space Sciences, University of Chinese Academy of Sciences, 19A Yuquan
Road, Beijing 100049, China\\
$^{3}$National Astronomical Observatories of China, Chinese Academy of Sciences, 20A Datun Road, 
Beijing 100020, China\\
}

\date{}
\pagerange{\pageref{firstpage}--\pageref{lastpage}} \pubyear{2020} 
\maketitle
\label{firstpage}

\maketitle

\begin{abstract} 
A puzzle in \NGC\, is how to secularly maintain the counter-rotating disc (CRD) from $0.2$ 
to $7\,$pc unambiguously detected by recent ALMA observations of molecular 
gas. Upon further dynamical analysis, we find that the Kelvin-Helmholtz 
(KH) instability (KHI) results in an unavoidable catastrophe for the disc developed at the 
interface between the reversely rotating parts. We demonstrate that a close binary of 
supermassive black holes provides tidal torques to prevent the disc from the KH catastrophe 
and are led to the conclusion that there is a binary black hole at the center of \NGC.
The binary is composed of black holes with a separation of $0.1\,$pc from GRAVITY/VLTI 
observations, a total mass of $1.3\times 10^{7}\sunm$ and a mass ratio of $\sim 0.3$ 
estimated from the angular momentum budget of the global system. The KHI gives rise to a 
gap without cold gas at the velocity interface which overlaps with the observed gap of hot 
and cold dust regions. Releases of kinetic energies from the KHI of the  disc 
are in agreement with observed emissions in radio and $\gamma$-rays. Such a binary is 
shrinking on a timescale much longer than the local Hubble time via gravitational waves, 
however, the KHI leads to an efficient annihilation of the orbital angular momentum and 
a speed up merge of the binary, providing a new mechanism for solving the long standing 
issue of ``final parsec problem''. Future observations of GRAVITY$^{+}$/VLTI are expected 
to be able to spatially resolve the CB-SMBHs suggested in this paper.
\end{abstract}
  

\section{Introduction}
For many decades, the Seyfert 2 galaxy \NGC\, served as an archetype for the unification 
model of active galactic nuclei  \citep{Antonucci1993} and has received the most intensive 
observations from radio, infrared, optical, ultraviolet, X-ray and $\gamma-$ray bands, 
which had been individually reported by several hundreds of papers (see Appendixes for some 
details). It is brighter than 
9.1\,mag with a redshift of $z=0.003793$ and a distance of 14.4\,Mpc). However, the nature 
of the \NGC\, nucleus, obscured by the dusty torus still remains unknown.

ALMA targeted CO\,($6-5$) molecular line and resolved a $7-10\,$pc geometrically thick-disc 
structure in \NGC. The dynamics shows non-circular motions 
and enhanced turbulence superposed on a surprisingly slow rotation pattern of the 
 disc \citep{Garcia-Burillo2016,Gallimore2016}. However, recent ALMA observations of HCN gas 
spatially resolved a counter-rotating disc (CRD) from 
$0.5-1.2\,$pc and beyond reversely extended to $7\,$pc
in the velocity map of HCN gas \citep{Imanishi2018,Impellizzeri2019}. In particularly, 
it has been found that the outer  disc shows Keplerian rotation consistent with extrapolation 
from the inner disc \citep{Impellizzeri2019}, providing invaluable information
of the CRD lifetime.

It has been realized that the CRD formation could be related somehow with potential fates 
of the two streamers at $\sim10\,$pc scale 
developed from the south and the north tongues by SINFONI/VLT \citep{Muller2009,Gravity2020}. 
Random collisions of molecular clouds from large scale distances
drive them into nuclear regions and trigger Seyfert activity \citep{Sanders1981}. 
Two tongues around the central black hole are formed by this way, and later could lead to 
formation of the CRD in \NGC. {\cblue As to the CRD one must explain how
it avoids being destroyed by the KHI on orbital time-scales.} 
Since this directly conflicts with the presence of the observed CRD, 
the mechanism that is secularly maintaining it at least as long as the viscosity timescale remains 
a puzzle.

\begin{figure*}\label{fig1}
\centering
\includegraphics[angle=0,width=0.825\textwidth,trim = 10 10 25 -15, clip]{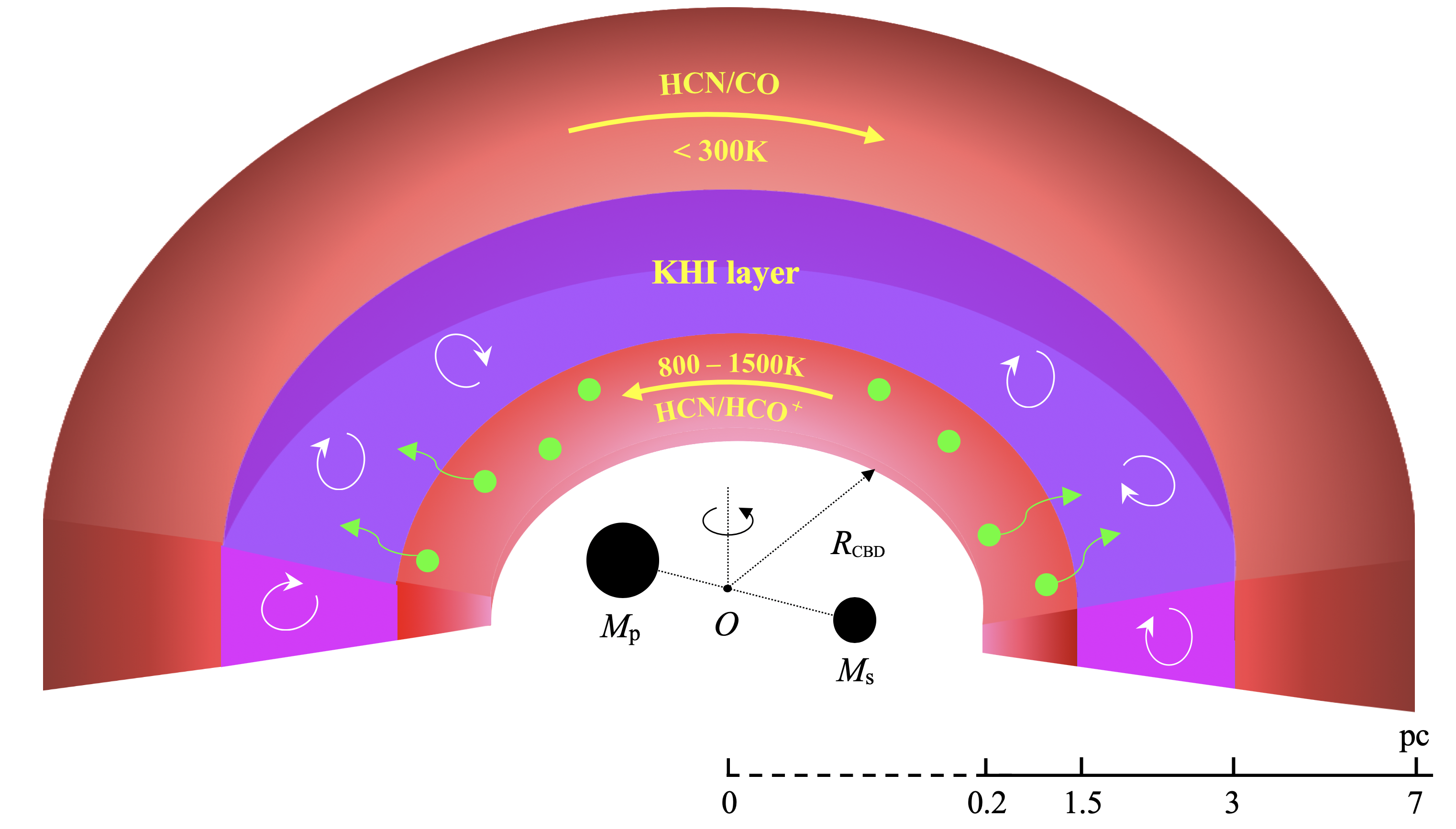}
\caption{\baselineskip 20.5pt
    { { Cartoon of a close binary of supermassive black holes
    maintaining one circumbinary  disc (CBD) composed of the prograde 
    ($R\lesssim 1.5\,$pc) and the retrograde ($3\,{\rm pc\lesssim {\it R}\lesssim 7\,pc}$) 
    parts in NGC\,1068 observed by ALMA \citep{Impellizzeri2019} and MIDI/VLTI \citep{Raban2009}.}} 
    {\cblue The heights measured by MIDI are $H_{0}/R\sim 0.3$ (inner part) and $\sim 0.6$ (outer
    part), respectively (see details in the text). $R_{\rm CBD}\sim 0.24\,$pc} is the inner edge 
    of the CBD measured by GRAVITY \citep{Gravity2020}, which 
    is consistent with dust sublimation radius. The prograde part is the NIR emission regions 
    where water maser clouds (green circles) are co-spaced. The retrograde part is the FIR emission 
    regions. The red color represents the dusty molecular  disc (HCN, HCO$^+$ and CO) radiating 
    in infrared. The interface regions in purple between the prograde and the retrograde parts are 
    undergoing the Kelvin-Helmholtz instability (KHI) to form shocks and turbulences, driving 
    formation of a gap with a width of $\deltaR\approx 0.82\calM H_{0}$. The KHI dissipates the 
    kinetic energy of the counter rotating  disc giving rise to bremsstrahlung emissions observable 
    in radio with morphology tightly related to the  disc shapes. Moreover, shocks in the KHI 
    layer accelerate some electrons 
    to be relativistic to radiate $\gamma$-ray emissions observed by {\it Fermi}-LAT.
        }
\end{figure*}

In this paper, we demonstrate that the CRD could be secularly maintained 
by a close binary of supermassive black holes (CB-SMBHs) supplying angular momentum to the CRD 
and avoiding the KH catastrophe. A model based on observations is provided for the binary system with 
a counter-rotating circumnuclear  disc. Such a system is showing evidence for an efficient 
way of removing the angular momentum through the retrograde accretion to harden the binaries.

\section{The circumnuclear  disc supported by binary black holes}
\subsection{Anatomy of the CBD in NGC 1068}
Figure 1 shows the geometric structure of the CRD obtained from ALMA, VLTI/MIDI, GRAVITY, VLBI 
and BLVI (see details in Appendix A). The CRD of HCN molecular gas in \NGC\, is composed of the 
inner counterclockwise (prograde: $0.5-1.2\,$pc) and the outer clockwise rotation 
(retrograde: $<\!7\,$pc) parts distinguished by the interface radius around 
$R_{*}\approx 2.5\,$pc revealed from the velocity map  \citep{Impellizzeri2019}. 

The black hole mass has been 
measured by water maser dynamics \citep{Greenhill1996,Greenhill1997}, but
corrected for some effects of massive  disc gravity \citep{Hure2002,Hure2011,Lodato2003}. 
However, mass of the disc within a few parsec is still dominated by the central black 
hole \citep{Imanishi2018}, questioning the validity of the corrections. Neglecting the uncertainties,
we take the black hole mass of $\sim 10^{7}\sunm$ in this paper, which should be robust. 
On the other hand, \NGC\, is a prototypical oval galaxy with an unusually massive 
pseudobulge \citep{Kormendy2013} of $M_{\rm bulge}=10^{10.9\pm0.1}\,\sunm$ and stellar dispersion 
velocity of $\sigma_{e}=151\pm7\,\kms$. Usually, pseudobulges are located below the well-known
$\bhm-\sigma$ relation \citep{Tremaine2002}, the upper limits of the black holes are 
$\bhm<3\times 10^{7}\sunm$ from this relation.

The KHI driven by shear collisions is unavoidable at interface regions, making the CRD 
lifetime just comparable with the Keplerian rotation timescale of 
$\tK=\Omega_{\rm K}^{-1}=1.3\times 10^4\,M_7^{-1/2}R_{\rm 2pc}^{3/2}\,${\rm yrs} 
through efficient annihilation of 
angular momentum (AM) simply estimated by the linear analysis \citep{Quach2015},
where $R_{\rm 2pc}=R/2\,{\rm pc}$ is the radius of  disc and $M_{7}=\bhm/10^{7}\sunm$ 
is the black hole mass. This is the KH catastrophe of the CRD in \NGC. For a thin  disc 
keeping a vertical equilibrium, it secularly evolves with the viscosity 
timescale \citep{Pringle1981,Pringle1991} of 
$t_{\rm vis}\approx \alpha^{-1}h^{-2}\tK
   =1.3\times 10^{7}\,\alpha_{0.1}^{-1}h_{0.1}^{-2}M_{7}^{-1/2}R_{\rm 2pc}^{3/2}\,$yrs,
where $\alpha=0.1\alpha_{0.1}$ is the viscosity parameter, $h=H_{0}/R=0.1\,h_{0.1}$ 
is the scale height of the  disc. Observations show that the  disc of HCN gas follows
Keplerian rotations in the both prograde and retrograde parts \citep{Impellizzeri2019}, 
indicating that the two parts are in dynamical equilibrium and therefore already exist 
at least as long as a lifetime of $t_{\rm vis}$. 

\subsection{Kelvin-Helmholtz instability}
The CRD is supersonically rotating. The Mach number of  disc rotation is 
$\calM=\VK/\cs^{0}\approx\left(H_{0}/R\right)^{-1}=10\,h_{0.1}^{-1}$ from $H_{0}=\cs^{0}/\Omega_{\rm K}$, 
where $\VK=R\Omega_{\rm K}$ is the rotation velocity and $\cs^{0}$ is the sound speed outside 
the KHI layer. In this paper, the subscripts of $\oplus$ and $\ominus$ symbols denote the 
prograde and retrograde rotations, respectively. In \NGC, GRAVITY observations show 
$h\lesssim 0.14$ in the dust sublimation ring, but $h\approx 0.3$ for the 
$R\lesssim 1.2\,$pc regions (near infrared regions with temperatures of $800-1500\,$K), namely,
$\calM_{\oplus}\sim 3$. In the mid-infrared regions ($\gtrsim 3\,$pc regions with temperatures 
$\lesssim 300\,$K) \citep{Raban2009}, 
we have $h\approx 0.6$, as a geometrically slim, and $\calM_{\ominus}\approx 1.7$.
Being different from the well-understood normal KHI modes \citep{Drazin2002}, 
both the prograde and the retrograde parts as supersonic flows lead to an extremely 
complicated interface from AM cancellations subject to black hole gravity and fast 
cooling of the hot plasma heated by the formed shocks. 
We take the simplest approach using the Rankine-Hugoniot conditions of shocks for the supersonic 
collisions of two streams in a slab geometry \citep{Lee1996}. Both the prograde ($v_{1}=v_{\rm K}$) 
and retrograde ($v_{2}=-v_{1}$) parts have the same Keplerian velocity $(v_{\rm K})$ but they 
have densities of $\rho_{0}$ and $\chi\rho_{0}$ ($\chi\le 1$) at the interface, respectively, 
where $\chi$ is the density ratio depending on properties of the CRD. The velocity, density 
and pressures of the post shock regions are $v/\VK=\left(1-\sqrt{\chi}\right)/(1+\sqrt{\chi})$,
$\rho/\rho_{0}=(\Gamma+1)/(\Gamma-1)$ and 
$p=2(1+\Gamma)\chi(1+\sqrt{\chi})^{-1}\rho_{0}\VK^{2}$, respectively, where $\Gamma$ is 
the adiabatic index. For the case with identical densities ($\chi=1$), we have the most 
efficient dissipation of the kinetic energy into the thermal since $v=0$. This gives 
rise to multiwavelength emissions tightly related to the dynamical structure of this region.

Supposing the interface radius $(\RKHI)$ with a thickness
of $\deltaR$, the AM cancellation driven by the KHI is happening with a timescale of 
$t_{\rm KHI}\approx \Omega_{\rm K}^{-1}$ and $\deltaR\approx t_{\rm KHI}\VR$, where $\VR$ 
is the velocity of gas entering the KHI zone. In a linear analysis,
$V_{\rm R}$ is given by the viscosity stress, which is much smaller than the sound speed 
($\cs$) in the KHI layer formed due to shear collisions. In the present context, however,
we have $\VR\approx c_{\rm s}$ in the layer because the efficient dissipation of the 
kinematic energy of the interface layer heats the medium and the waves propagate with 
$\cs$. With the shock conditions, we have 
$c_{\rm s}^{2}=p/\rho=2\chi(1+\sqrt{\chi})^{-1}(\Gamma-1)\VK^{2}=(2/3)\VK^{2}$,
for $\chi=1$ and $\Gamma=5/3$, 
which results in the KHI gas to have two thirds of the virial temperatures.
Considering $H_{0}=\cs^{0}\Omega_{\rm K}^{-1}$, we have the layer width  
approximated by 
$\deltaR\approx t_{\rm KHI}\cs\approx 0.82\,\Omega_{\rm K}^{-1}\VK=0.82\,\calM H_{0}$.
For the current case of 
\NGC, the interface of the density map is not resolved by the current ALMA 
observations \citep{Impellizzeri2019}, but the velocity interface is quite clear from 
the velocity map (see Figure 2 in \cite{Impellizzeri2019}). The projected radius 
is about $R_{*}\approx 2.5\,$pc, which exactly overlaps with the geometrical gap between 
the NIR and MIR component \citep{Raban2009}. We thus take the interface radius 
$R_{\rm KHI}=R_{*}$. For the simplest estimation, we just take $\calM_{\rm KHI}\approx 2.5$ 
before shear collisions, which is the averaged value between the NIR and MIR regions. 
The KHI layer therefore has a width of $\deltaR\approx 2\,H_{0}\approx 1.6\,$pc, where 
$H_{0}\approx 0.8\,$pc is the averaged height of the NIR and MIR regions in light 
of observations \citep{Raban2009}. This gap should be observed in density map, however,
it is not spatially resolved ($1.4\,$pc for \NGC) of the current ALMA \citep{Impellizzeri2019}.
However, we note that the width of the gap 
is affected by the prograde part supported by the tidal torques of one CB-SMBH. 

The KH catastrophe can be evaluated by AM distributions of the CRD.
The total AM of the retrograde part (within 7\,pc) can be estimated by 
$\calL_{\ominus}\approx -M_{\rm gas}^{\ominus}v_{\rm K}\bar{R}_{\ominus}
                \approx -325.3\,M_{7}^{1/2}\,\sunm{\rm\,pc^{2}\,yr^{-1}}$,
where $\bar{R}_{\ominus}=5\,$pc is the averaged radius of the retrograde regions (3-7\,pc),
and $v_{\rm K}\approx 10^{7}\,{\rm cm\,s^{-1}}$ at this radius. For the prograde part ($<\!1.5\,$pc),
we have the total AM of $\calL_{\oplus}\approx 190.9\,M_{7}^{1/2}\sunm{\rm\,pc^{2}\,yr^{-1}}$, where
we take $M_{\rm gas}^{\oplus}=9\times10^{5}\,\sunm$ and the averaged radius 
$\bar{R}_{\oplus}=1\,$pc. Owing to $|\calL_{\oplus}|<|\calL_{\ominus}|$, the prograde part 
will be destroyed by the retrograde via the KHI action.
Given an AM distribution of a  disc as $\calL\propto R^{\beta}$ ($\beta>0$), the cancelled 
AM can be estimated by $\delta\calL/\calL=\beta\left(\deltaR/\RKHI\right)$. From  
$\deltaR\approx 0.82\calM H_{0}$, we have $\delta\calL/\calL\approx 0.82\beta$. Though 
the gaseous dynamics has been measured in \NGC, the spatial distribution of mass density 
is poorly done. In the ${\rm 14\,pc\times 10\,pc}$ regions \citep{Imanishi2018}, the mass is 
about $2\times10^{6}\sunm$ corresponding to an averaged surface density of 
$\Sigma_{\rm gas}^{\ominus}\approx 1.4\times 10^{4}\sunm{\rm pc^{-2}}$ over this region. We have
$\Sigma_{\rm gas}\propto R^{-0.6}$ from the prograde ($\bar{R}_{\oplus},\Sigma_{\oplus}$ given in
SM) to retrograde ($\bar{R}_{\ominus},\Sigma_{\ominus}$) parts, 
giving rise to $\calL\propto R^{1.9}$. 
With $\beta\approx 1.9$, we find that the AM of the CRD can be cancelled 
completely by once action of the KHI within a dynamical timescale of 
$t_{\rm KHI}\sim \Omega_{\rm K}^{-1}
            =1.9\times 10^{4}\,M_{7}^{-1/2}\left(R_{\rm KHI}/R_{*}\right)^{3/2}\,$yrs. 
This is a very short timescale compared with the typical AGN lifetime, even the 
flickering lifetime ($\sim 10^{5}\,$yrs) \citep{Schawinski2015}. It would be worse for 
the prograde part if the retrograde part is able to obtain more AM from the outer boundary. 
The KH catastrophe cannot be avoided unless there is an extra
supplement from external sources of torque. Obviously, the only way to maintain the prograde 
part is the gain AM from a CB-SMBH.

\subsection{Close binary of supermassive black holes}
As seen in a brief review in Appendix B, evidence for a CB-SMBH is still quite elusive 
in observations though there are a few candidates.
Starting from one CB-SMBH with a circular orbit of a separation ($A$) composed of
the primary $(M_{\rm p})$ and the secondary $(M_{\rm s})$, we have its period of
$P_{\rm orb}\approx 877\,(1+q)^{-1/2}M_{7}a_{5}^{3/2}\,{\rm yrs},$
and orbital AM of 
$\calL_{\rm B}=656.2\,q(1+q)^{-1/2}M_{7}^{2}a_{5}^{1/2}\,\sunm{\rm\,pc^{2}\,yr^{-1}}$,
where $q=M_{\rm s}/M_{\rm p}$ is the mass ratio, $a=A/\Rs$, $a_{5}=a/10^{5}$,
$\Rs=2GM_{\rm p}/c^{2}=3.0\times 10^{12}\,M_{7}\,$cm is the Schwarzschild radius, $G$ is 
the gravitational constant and $c$ is the speed of light. The tidal torques of one CB-SMBH 
strongly interact with its circumbinary  disc (CBD) to form a cavity with an inner radius 
of $R_{\rm CBD}=2A$ in light of analytical calculations and 
simulations \citep{Artymowicz1994,Armitage2002}. In Appendix C, we demonstrate that 
2$\mu$m-imaged hole discovered by GRAVITY \citep{Gravity2020} is approximate to 
$R_{\rm CBD}=0.24\,$pc with help of the gas dynamics of ALMA-detected CRD. The timescale 
of gravitational waves is given by $t_{\rm GW}=2.0\times 10^{14}\,M_{7}a_{5}^{4}/q(1+q)\,$yrs 
for circular orbits \citep{Peters1964} much longer than the Hubble time, 
which is the well-known ``final parsec problem". To maintain the prograde part, we need 
$\calL_{\rm B}\gtrsim \calL_{\oplus}$, yielding $q\gtrsim 0.3$ (for $M_7=a_{5}=1$) in order 
to make up for the AM loss due to the KHI action. 

The CBD regions governed by the tidal 
torques of the CB-SMBH can be estimated in a simple way. Tidal forces can be approximated 
by $F_{\rm tid}\approx \left({GM_{\rm tot}\delta m}/{R^{2}}\right)\left({A}/{R}\right)$ for 
a point mass of $\delta m$, which is about perpendicular to the direction to the center, 
and the corresponding torques are
$\calT\approx F_{\rm tid}R=\delta m (1+q)c^{2}ar^{-2}/4$,
where $r=R/\Rs$. {\cblue Taking the ratio of the gas 
parcels angular momentum ($\Delta\calL$) and the binary torque ($\Delta\calT$), 
we define the tidal timescale
of $t_{\rm tid}=\Delta \calL/\calT$, where}
$\Delta\calL=\delta m(1+q)^{1/2}\Rs cr^{1/2}/\sqrt{2}$, 
yielding 
$t_{\rm tid}=(1+q)^{-1/2}(a/r)^{-1}\tK$.
Considering the rapid decays of the tidal torques with radius \citep{Papaloizou1984,Pringle1991}, 
we recast the tidal timescales of
$t_{\rm tid}=\varepsilon_{0}(1+q)^{-1/2}(a/r)^{-4}\tK$, where 
$\varepsilon_{0}\approx 0.1$. In the tidal-governed regions $t_{\rm tid}\le t_{\rm vis}$, 
we have the critical radius of 
$R_{\rm tid}\le 10\,\left(\varepsilon_{0.1}\alpha_{0.1}\right)^{-1/4}h_{0.1}^{-1/2}(1+q)^{1/8}\,A$, 
where $\varepsilon_{0.1}=\varepsilon_{0}/0.1$, beyond which the viscosity torques becomes 
dominant. For \NGC, we have $R_{\rm tid}\approx 1.2\,$pc for typical values 
$(\varepsilon_{0.1},\alpha_{0.1},h_{0.1})=1$, which is in agreement with the sizes 
of the prograde part \citep{Impellizzeri2019}. Such a CB-SMBH guarantees the stability 
of the prograde part and prevents it from the KH catastrophe.

The gas  disc beyond $R_{\rm tid}$ is jointly governed by viscosity stress and the tidal 
torques, and its outer boundary conditions. Since $R>R_{\rm tid}$ part is retrograde 
with respect to the CB-SMBH orbits, it loses AM due to the tidal torques, enhancing 
accretion onto the center \citep{Nixon2011,Roedig2014,Bankert2015}. In the context of the 
inner part as a decretion  disc \citep{Pringle1991}, its outer boundary moves outward with 
the viscosity timescale but encounters the inward flows from the retrograde part, leading 
to an equilibrium radius ($\RKHI$) between the inward and the outward flows. Dynamical
evolution of the CRD could compress the width of the gap. Numerical 
simulations are expected for details of such a complicated case. For \NGC, 
$\calL_{\rm B}>|\calL_{\oplus}+\calL_{\ominus}|$ indicates that 
the CB-SMBH with a total mass of $1.3\times 10^{7}\sunm$, mass ratio of $q=0.3$ and 
a separation of $0.1\,$pc can support this observed CRD in the galactic center. 

On the other hand, the CB-SMBH in \NGC, as the first observed object,  
faces the well-known ``final parsec problem" \citep{Begelman1980,Goicovic2018}. 
{\cblue With the orbital parameters obtained presently, we can 
estimate the orbital evolution purely driven by gravitational waves. 
According to \cite{Peters1964}, the shrinking timescale of orbits is given by
$t_{\rm GW}=1.24\times 10^{13}\,a_{5}^{4}M_{7}/q(1+q)\,$yrs, which is much longer 
than the Hubble time and episodic lifetime of AGNs. 
This indicates that the CB-SMBHs can not merger via radiations 
of GWs within the Hubble time, showing ``the final parsec problem''.}
Successive series of random accretion of molecular clouds onto binary black holes
have been suggested by simulations \citep{Goicovic2018}, fortunately, from
observational side, two tongues
from south and north direction around the center appear in \NGC\, $\sim10\,$pc 
regions \citep{Muller2009}. 
The retrograde part can efficiently cancel the binary orbital AM through the KHI of
random and episodic accretion. The episodic numbers can be roughly estimated by 
$N=\calL_{\rm B}/|\calL_{\ominus}+\calL_{\oplus}|$, and accretion with low AM 
(i.e., $|\calL_{\ominus}+\calL_{\oplus}|\approx 0$) can 
efficiently shrink the CB-SMBH in light of $t_{\rm GW}\propto M_{\rm p}^{-3}$ for
the case given $A$. For \NGC, we have $N\approx 4.8$ 
from ALMA observations \citep{Imanishi2018,Impellizzeri2019}. This indicates that 
the retrograde accretion onto CB-SMBHs is an efficient way of removing orbital
AM of the binaries.

\subsection{Multi-$\lambda$ emissions}
During the AM cancellation through the KHI, a huge amount of kinematic energies of the CRD 
is released, which can be estimated by $\Delta E_{\rm K}\approx \Delta M_{\rm KHI}\VK^{2}$ 
within the interval of $\Omega_{\rm K}^{-1}$. The radiative luminosity from the dissipation 
is given by 
$L_{\rm KHI}\approx \left(\Delta M_{\rm KHI}/\Delta t_{\rm KHI}\right)\VK^{2}
            =\ell_{*}\dot{M}_{\rm torus}c^{2}$,
where $\ell_{*}=1/(r_*r_{_{\rm KHI}}\alpha h^{2})=2\times 10^{-4}\,\alpha_{0.1}^{-1}h_{0.1}^{-2}$ 
(taking $r_{_{\rm KHI}}=1$ and $r_*=R_*/\Rs=5\times 10^6$) 
and $\dot{M}_{\rm torus}$ is the accretion rates of the torus.
For the typical values of the parameters, we have 
$L_{\rm KHI}\approx 1.1\times 10^{44}\,\dot{M}_{1}\ergs$ at the peak frequency of 
$\epsilon_{\rm peak}\sim 0.1\,r_{\rm KHI}^{-1}\,$keV from the hot KHI layer during the interval 
of the KHI timescale, where $\dot{M}_{1}=\dot{M}_{\rm torus}/10\,\sunm\,{\rm yr^{-1}}$. The 
infall with a typical rate of $10\,\sunm\,{\rm yr^{-1}}$ at
a few parsec scales in \NGC\, has been observed by SINFONI \citep{Muller2009,Garcia-Burillo2019} 
(most of this rate will be channeled into outflows). Such a powerful emission can be detectable 
in type 1 AGNs, however, for \NGC\, as a Compton thick AGN \citep{Zaino2020}, the major emissions 
from the KHI layer will be undetectable because of absorption of the torus along the line of 
sight. The KHI provides a new explanation of soft X-ray excess if there is a CRD like in \NGC.  
The excess should be relatively stationary because of its origins from a large scale.

On the other hand, fortunately, radio emission from this HKI dissipation is detectable,
which can be estimated by the bremsstrahlung emission from the KHI layer. 
The shear collisions of the two streamers make the post-shocked gas with temperatures 
close to the virial one as $T_{\rm gas}\approx 1.2\times 10^{6}\,r_{_{\rm KHI}}^{-1}$K, 
where $r_{_{\rm KHI}}=\RKHI/R_{*}$. Since the bremsstrahlung cooling timescale of 
hot plasma is \citep{Rybicki1979} $t_{\rm cool}\approx 3.0\,n_{6}^{-1}T_{6}^{1/2}$yrs, 
the KHI layer can be efficiently cooled, where $n_{6}=n_{e}/10^{6}\,{\rm cm^{-3}}$ is the 
density (about the CRD gas density) and $T_{6}=T/10^{6}\,{\rm K}$ is the layer temperature. 
The fast cooling ($t_{\rm cool}\ll \Omega_{\rm K}^{-1}$) prevents the shocks 
from propagation outside the layer. Owing to the black hole gravity, the 
cooled layer will rapidly collapse onto the prograde part with the free-fall timescale of 
$t_{\rm ff}\approx 1.8\times 10^{4}\,r_{\rm KHI}^{3/2}\,$yrs. On the other hand, gas is 
supplied beyond the KHI layer with a timescale of 
$t_{\rm vis}\approx 10^{3}\,\alpha_{0.1}^{-1}h_{0.1}^{-2}\tK$ 
by the viscous stress from the retrograde part. The comparison of 
$t_{\rm cool}\ll t_{\rm KHI}\sim t_{\rm ff}\ll t_{\rm vis}$ implies that the gas supply 
to the layer will be halted. The cooled gas in the layer will be emptied by the gravity 
of the central black holes, forming a gap with a width of $\deltaR$. 
Fortunately, such a gap has been revealed by MIDI/VLTI, showing spatial separations 
between the NIR (hot dust with temperature $>800\,$K, and located $<1.4\,$pc) and MIR 
($<300\,$K, and located $>3\,$pc) emissions \citep{Raban2009}.

With the KHI layer's  
volume of $V_{\rm KHI}=2\pi R_{\rm KHI}\deltaR H_0=1.6\pi \RKHI^{3}\calM\left(H_0/R\right)^{2}$,
we have radio emissions of 
$L_{256\rm GHz}=1.0\times 10^{39}\,n_{6}^{2}T_{6}^{-1/2}R_{2.5}^{3}h_{0.1}^{2}\,\ergs$ 
at $\nu=256\,$GHz
according to the thermal emissivity [Eq.5.14b in \cite{Rybicki1979}]. This radio 
emission agrees with $12.7\,$mJy observed by ALMA  \citep{Impellizzeri2019} at $256\,$GHz. 
As emitted from the layer, the radio morphology should follow the velocity interface. 
Indeed, this agrees with the observations by comparing the radio 256\,GHz map with the HCN 
velocity map from Figure 2 in \cite{Impellizzeri2019}, showing the north-east part 
of the HCN  disc and southwest part where the velocity interfaces matches the radio morphology. 
Moreover, the layer has a luminosity of 
$L_{\rm 5\,GHz}\approx 2.0\times 10^{37}\,\ergs$ at 5GHz from bremsstrahlung emissions, 
which roughly agrees with the compact radio source $S_{1}$ in the nuclear center.

Additionally, the strong shocks unavoidably accelerate some electrons to be relativistic in the 
KHI layer, leading to non-thermal high energy emissions from the layer. {\it Fermi}-LAT observations
 \citep{Ajello2017} detected a $\gamma$-ray luminosity of about $2.5\times 10^{41}\,\ergs$ with 
a cut-off energy of $\sim 10^{2}\,$GeV detected by MAGIC \citep{Acciari2019}. Jet model and starburst 
model are suggested, but origins of the GeV emissions remain open \citep{Acciari2019} [starburst model 
is not able to produce the observed GeV emissions, see Figure 2 in \cite{Acciari2019}].
Here the dissipation of the KHI layer could provide an alternative
explanation. Magnetic fields are of $B=10^{-3}\,$Gauss in this region \citep{Krolik1987}. The shock
velocity is $V_{\rm sh}\sim 3\times 10^{7}\,T_{6}^{1/2}\,{\rm cm\,s^{-1}}$, if $\Gamma=5/3$, 
yielding the acceleration
timescale of $t_{\rm acc}=R_{\rm L}c/V_{\rm sh}^{2}=5.8\times 10^{6}\,\gamma_{5}B_{-3}^{-1}T_{6}\,$sec,
where $\gamma_{5}=\gamma/10^{5}$ is the Lorentz factor of relativistic electrons,
$R_{\rm L}$ is the Lamore gyration radius \citep{Blandford1987}. The energy density of infrared photons
is about $U_{\rm IR}=1.8\times 10^{-6}\,L_{43}R_{\rm pc}^{-2}\,{\rm ergs\,cm^{-3}}$ dominating over the
magnetic fields, where $L_{43}=L_{\rm IR}/10^{43}\,\ergs$ is the infrared luminosity and 
$R_{\rm pc}=R/1\,{\rm pc}$ is the distance of the accelerating regions to the center. 
In such a context, the cooling of the electrons is governed by inverse Compton
scatter and given by
$t_{\rm IC}=3m_{e}c/4\sigma_{\rm T}\gamma U_{\rm IR}
           =1.7\times 10^{8}\,\gamma_{5}^{-1}L_{43}^{-1}R_{\rm pc}^{2}\,$sec  \citep{Rybicki1979},
and the relativistic electrons contribute less in radio band. The maximum Lorentz factor is 
$\gamma_{\rm max}=5.5\times 10^{5}\,L_{43}^{-1/2}R_{\rm pc}B_{-3}^{1/2}T_{6}^{-1/2}$ 
from $t_{\rm IC}=t_{\rm acc}$. The maximum energetic photons are 
$\epsilon_{\rm max}\approx \gamma_{\rm max}^{2}\epsilon_{\rm IR}
                    =74\,(\gamma_{\rm max}/\gamma_{0})^{2}\epsilon_{0.25}\,$GeV, 
where $\gamma_{0}=5.5\times 10^{5}$, $\epsilon_{0.25}=\epsilon_{\rm IR}/0.25\,{\rm eV}$ is 
IR photon energy in units of $\lambda=5\mu$m estimated from the spectral energy 
distribution \citep{Gravity2020}. If the total budget of $\gamma$-rays is conservatively 
1\% [see \cite{Blandford1987}] of $L_{\rm KHI}$, about $10^{42}\ergs$ 
is produced that is enough to explain the observed $\gamma$-rays.

\section{Discussions}
\subsection{Magnetic fields}
We first consider the role of a strong azimuthal magnetic field to support the CRD. 
The Alfv\'en velocity is given by $v_{\rm A}=B_{\phi}/\sqrt{\mu_{0}\rho_{0}}$, where $B_{\phi}$
is the azimuthal component of the magnetic field, 
$\mu_{0}$ is permeability of free space and $\rho_{0}$ is the density
of gas. The plasma couples with neutral part tightly so that the magnetic fields controlling 
plasma govern the neutral part of gas. If the azimuthal magnetic field supports the CRD, the 
Alfv\'en waves provide enough energy to secularly support the CRD, namely 
$v_{\rm A}^{2}\ge \left(t_{\rm vis}/t_{\rm K}\right) \left(\Delta V\right)^{2}$, where 
$\Delta V\approx 2v_{\rm K}$ is the differential velocity of the CRD. The magnetic field is 
given by
\begin{equation}
B_{\phi}\ge 662\,\alpha_{0.1}^{-1/2}h_{0.1}^{-1}\rho_{-18}^{1/2}v_{2}\,{\rm mG},
\end{equation}
where $\rho_{-18}=\rho_{0}/10^{-18}\,{\rm g\,cm^{-3}}$ [the ALMA-observed density of gas from 
the HCN line \citep{Imanishi2018}] and $v_{2}=v_{\rm K}/10^{2}\,{\rm km\,s^{-1}}$. 
Recent polarization observations show $B\approx 0.7\,$mG 
in the few parsec regions of its circumnuclear region  \citep{Lopez2020} in \NGC, which is much lower
than the lower limit of the magnetic fields to support the CRD. This directly rules out the
magnetic fields to support the unusual CRD. 

\subsection{A CRD around a single SMBH}
Since nuclear regions are much smaller than the circumnuclear disk, random accretion onto black 
holes is naturally taking place \citep{Sanders1981,King2006}. In a cosmic timescale, such 
a kind of the random accretion leads to spin-down evolution of SMBHs as a result of 
cancellations of spin AM from different episodes \citep{Kuznetsov1999,King2008}, but speeds 
up cosmic 
growth of SMBHs due to low radiation efficiency. According to the cosmic evolution of
duty cycle of SMBH activity episodes \citep{Wang2006}, the spin-down behaviors of SMBH 
evolution are found from application of cosmic equation of radiative efficiency to 
the survey data \citep{Wang2009,Li2012}, providing strong evidence for the random accretion. 
In particular, cold clumps are universal in galaxies \citep{Tremblay2016,Temi2018}. 
On the other hand, this indicates that there were a series 
of CRDs across cosmic time. The CRD will give rise to very fast accretion onto SMBHs [also 
see Ref\, \citep{Kuznetsov1999,Quach2015,Dyda2015}] and show some imprints
at scale of a few parsec. Is there any independent 
evidence for this?

Thanks are given to recent ALMA observations of nearby Seyfert galaxies
for a new clue to understanding this issue. Among this sample, three Seyfert galaxies 
(NGC\,1365, NGC\,1566 and NGC\,1672) show a gas hole or depletion in the center within 
a few parsec \citep{Combes2019}, of which two are more nearby than \NGC. 
It is very interesting to note that NGC\,1365 and NGC\,1566 are 
changing-look AGNs. NGC\,1365 is changing its column density from Compton-thick to 
-thin \citep{Winter2009} whereas NGC\,1566 is changing spectral type \citep{Oknyansky2019}.
We preliminarily speculate that the gas hole could be caused by the KHI, which leads to 
AM annihilation of CRDs depleting the parsec scale regions. If this happened, a fast 
collapse onto very central regions results in high accretion rates of AGNs. The parsec 
scale hole of gas could be an indicator of the remanent of a past CRD around a single 
SMBH. Considering the short lifetime of a CRD around single SMBH as low as 
$t_{\rm KHI}/t_{\rm vis}\sim 10^{-3}$, we think that the CRD will be detected in an 
extremely low opportunity. Future MUSE observations of ALMA-observed targets 
should explore detailed structure at a few parsec scales to test if there are something 
related to ``two tongues'' in \NGC\, \citep{Muller2009}. Additionally, we should check their 
polarized spectra for optical \feii\, emission lines for accretion status of the central 
black holes \citep{Du2019}. On the other hand, candidate AGNs harboring CB-SMBHs with 
CRDs can be selected from Seyfert 2 galaxies whose polarized spectra show asymmetric 
profiles of broad emission lines.

{\cblue \subsection{Disc geometry}
For simplicity, we apply the Rankine-Hugoniot calculation to the slab geometry of the CRD
in this paper, allowing us to focus on analysis of the major effects of the KHI action and 
tidal interaction of the CB-SMBHs with its circumbinary disk
in a simple way. Following \cite{Quach2015} and \cite{Dyda2015}, we will 
make use of disc geometry of the counter-rotating gas for detailed analysis of the KHI and 
its effects in the future. The violent dissipation of the kinetic energies of the
counter-rotating motion gives rise to gas expansion in the vertical direction so that we have
to study 3-dimensional processes, such as outflows from the disc surface, which
could be probably related with the observed in this region.  
All these are beyond the scope of this paper, but we will carry out detailed applications of
the KHI theory to \NGC\, in an separate paper.
}

\section{Conclusions and Remarks}
The puzzle of a counter-rotating disc from $0.2-7$\,parsec regions discovered by ALMA 
observations arises a serious issue how to secularly maintain it. The Kelvin-Helmholtz
instability will destroy the disc unless there is an extra supply of angular momentum.
A close binary of supermassive black holes is expected to support the unusual disc for
a viscosity timescale. The binary black holes are of a total mass of $1.3\times 10^{7}\sunm$
and mass ratio of $\gtrsim 0.3$ and a separation of 0.1\,parsec. With annihilation of
angular momentum due to the KHI, the disc efficiently dissipates its kinematic energy
into radiation peaking at soft X-rays and radio bands from bremsstrahlung emissions.
Though soft X-rays are absorbed by the torus, the radio emissions are in agreement with
observations. We also find significant $\gamma$-rays from the shocks developed from the 
KHI, which agrees with {\it Fermi} and MAGIC observations. 

The polarized spectra of \NGC\, show asymmetric profiles of broad
H$\beta$ (with polarization degree of $\sim 3\%$) \citep{Miller1991} indicating 
complicated structure of the broad-line regions. In the NIR, the core has a 
$m_{K}\approx 8.7\,$mag \citep{Gravity2020}, the polarized magnitudes 
could be of $\sim 12.5\,$mag in $K-$band. The polarized fluxes of broad Brackett$\gamma$ 
line \citep{Martins2010} could be too weak to detect through the GRAVITY/VLTI. 
Fortunately, it is strong enough for GRAVITY$^{+}$ as the next generation of GRAVITY/VLTI 
to make efforts to spatially resolve the $<0.1\,$ parsec regions for the appearance
of the CB-SMBHs through differential phase curves \citep{Songsheng2019}. 

Given the black hole mass, mass ratio and separations, the CB-SMBH is radiating gravitational 
waves (GW) with frequencies of $f_{\rm GW}=0.1\,(1+q)^{1/2}M_{7}^{-1}a_{5}^{-3/2}\,$nHz and 
intrinsic strain of amplitudes is $h_{\rm s}=5\times 10^{-19}\,qM_{7}a_{5}^{-1}d_{14.4}^{-1}$, 
where $d_{14.4}=d/14.4\,{\rm Mpc}$ is the distance of \NGC\, to observers. 
GW background composed of NGC 1068-like AGNs can be detected by Square Kilometer 
Array (SKA) \citep{Barack2018}. However, the retrograde part eventually shrink the 
CB-SMBH through cancelling its orbital AM to merge, providing a paradigm of
solving the ``final parsec problem", which is related with many concomitant phenomenon 
risen from the KHI. The CB-SMBH in \NGC, as one of long-sought-after 
candidates, has an angular size of $2\,$mas, which is larger by a factor of 100\, than 
the spatial resolution of the Event Horizon Telescopes (EHT), and is thus worth observing 
via EHT. 

\section*{Acknowledgements}
{{\cblue The authors are grateful to S. Dyda as the referee for a helpful report improving 
the manuscript.} Yao Chen is acknowledged for helpful discussions. Bo-Wei Jiang is thanked
for the plot in this paper.
JMW acknowledges financial support from the the National 
Science Foundation of China (11833008 and 11991054), from the National Key R\&D Program 
of China (2016YFA0400701), from the Key Research Program of Frontier Sciences of the 
Chinese Academy of Sciences (CAS; QYZDJ-SSW-SLH007), and from the CAS Key Research 
Program (KJZD-EW-M06).}

\appendix

\section{Observational properties}
{\it Polarized spectra.}
The polarized spectra of \NGC\, show typical spectra of Seyfert 1 
galaxies \citep{Antonucci1985,Miller1991} and 
hence demonstrate the existence of an obscured broad-line region (BLR) by a dusty torus. In light 
of the scenario, an orientation-based 
unification scheme was suggested for all Seyfert galaxies as well as for quasars \citep{Antonucci1993}. 
Broad H$\beta$ line appears in the optical polarized spectra along with strong and broad 
\feii\, emission lines.  Using the polarized spectra, the black hole
mass can be estimated by the virialized motion of BLR. We find the black hole mass
of $(8.4\pm 0.4)\times 10^{6}\sunm$ from the $5100\,$\AA\, luminosity estimated from the 
\oiii\, line \citep{Du2017},
which is consistent with results estimated from 2-10\,keV luminosity \citep{Zaino2020}.

It is important to note that the polarized H$\beta$ profile is asymmetric \citep{Miller1991}, 
implying complicated structures of the hidden BLR, at least existence of sub-structures of 
ionized gas \citep{Du2018}. Since 
the polarized lights to observers are from scatters of an electron screen vertically located at a 
distance of 100\,pc from the galactic center, it will be very hard to observe variations of the 
polarized H$\beta$ line through a feasible length campaign. Actually, the Brackett $\gamma$ line 
profiles are double-peaked, see Figure 4 in \cite{Martins2010}. \NGC\, is bright enough in 
near infrared for for GRAVITY spectroastrometry to measure the differential phase curves. Signatures 
of CB-SMBHs could appear in the differential phase curves \citep{Songsheng2019}.

{\it Dusty  discs.} 
The most important information for infrared continuum is the extinction, which can be estimated by 
near infrared emission lines. The Brackett $\alpha$ broad line component observed by Infrared Space 
Observatory (ISO) constrains a lower limit of the extinction $A_{\rm Br\alpha,4.05}\ge 2.4$, implying 
the extinction $A_{\rm K}\ge 6$ in the $K$-band \citep{Lutz2000}.
Observations of mid-infrared interferometric instrument (MIDI) for \NGC\, 
reveal two spatially resolved regions: 1) inner and hot ($>800\,$K) zone with 1.35\,pc long and 
0.45\,pc thick at a ${\rm PA}=-42^{\circ}$, and 2) outer and warm ($\sim 300\,$K) zone with 
$3\times 4\,$pc \citep{Jaffe2004,Raban2009} but extended to 7\,pc regions along the north-south 
axis \citep{Lopez2014}, and somehow may correspond to polar dust emission in the ionization cone.  
The gap between the NIR and MIR dust regions can be approximated by $R_{*}\approx 2.5\,$pc, actually
overlap with the velocity interface of the HCN gas map \citep{Impellizzeri2019}. We thus adopt it
as the interface radius where the KHI is taking place. 

GRAVITY onboard Very Large Telescope Interferometer (VLTI) 
observations find the dust sublimation radius $R_{\rm sub}=0.24\,$pc, where hot dust
particles ($\sim 1500\,$K) are responsible for NIR and MIR emissions with extinctions of 
foreground dust ($A_{\rm K}=5.5$) \citep{Gravity2020}. A structure with a hole has been found 
around $0.24\,$pc \citep{Gravity2020}. This radius agrees with the empirical relation established 
by NIR dust reverberation mapping of AGNs \citep{Minezaki2019}. However, the innermost part of the 
dusty  disc is geometrically thin as $H/R\lesssim 0.1$, which is still much thicker than the 
Shakura-Sunyaev  disc. 

{\it Maser  disc.} Water (and OH) masers observed by VLBI trace inner edge of the torus 
from a radius of $0.4$ to $\sim 1\,$pc with a ${\rm PA}=-45^{\circ}$ completely different from radio 
jet \citep{Greenhill1996,Greenhill1997,Gallimore1996b,Gallimore1996c}. The inclination of the maser 
 disc is $\sim 80^{\circ}$ whereas an inclination of $\sim 70^{\circ}$ is obtained by fitting infrared
continuum \citep{Lopez2018}. Generally the maser  disc aligns with the dusty  disc though slightly 
orientated differently.

{\it Radio morphology.}
A radio core source, called $S_{1}$, associated with the  disc of ${\rm H_{2}O}$ vapor megamaser 
emission has been found \citep{Greenhill1996,Gallimore2001}. It is interesting to find that $S_{1}$ 
overlaps with the inner prograde HCN  disc \citep{Impellizzeri2019}.   ALMA observations of $256\,$GHz 
continuum show very interesting features, not  discussed in Ref\, \citep{Impellizzeri2019}, that the 
radio flux contours overlap with the velocity interface between the two counter rotating parts of 
the HCN  disc, see Figure 2 in Ref\, \citep{Impellizzeri2019}.

\section{Close binary of supermassive black holes} 
As expected for several decades from the theory of the merger tree of galaxy evolution, 
CB-SMBHs (defined as those with separations 
less than 0.1\,pc) evolved from dual galactic cores must be located in centers of some 
galaxies sometimes \citep{Begelman1980,Volonteri2003}. Most phenomenon of AGN activities
can be explained by accretion onto a single SMBH \citep{Rees1984,Osterbrock1986}, 
implying that most CB-SMBHs have finished
the final mergers. Single epoch spectrum with double-peaked profiles cannot be used 
as a direct diagnostic of CB-SMBHs \citep{Popovic2012,Gaskell2010}. Though dual AGNs 
are quite common among galaxies \citep{Comerford2013,Fu2015,Liu2018}, however, observational 
evidence for CB-SMBHs is very elusive \citep{Popovic2012,Dotti2012}. Though the radio map of 
NGC 7624 
($z=0.0029$) shows two radio cores with a separation of 0.35 pc in the nucleus and plausibly 
implies a candidate binary black holes \citep{Kharb2017}, none of CB-SMBHs are identified so 
far. A very brief summary is given below about the 
progress of searching for CB-SMBHs. 

{\it The final parsec problem.}
The formation of massive black hole binaries is unavoidable as a outcome of successive 
mergers of galaxies, however, it turns out that the decay of the binary 
orbits in a real galaxy would be expected to stall at separations of parsec scales unless some 
additional mechanism is able to efficiently remove the orbital AM  \citep{Milos2001}. At large
separations, dynamical friction of the binaries with background stars controls orbital 
evolution  \citep{Milos2001,Yu2002,Wang2012}, but becomes inefficient when the mass enclosed in
the orbits are comparable in gas-poor environment and hardly evolve into stages of radiating 
gravitational waves. This is the so-called ``final parsec problem"  \citep{Begelman1980,Milos2001}. 

Numerous simulations show that the gas-rich environment can efficiently absorb and 
transfer the orbital AM through the tidal interaction with the prograde 
CBD  \citep{Escala2005,MacFadyen2008,Cuadra2009,Lodato2009}, or the 
retrograde  \citep{Nixon2011,Roedig2014,Bankert2015}. However, the real situations are 
complicated by successive random accretion of clumpy clouds formed in circumnuclear 
regions  \citep{Tremblay2016,Temi2018}, which have been found as an efficient clue to harden
binary black holes from simulations  \citep{Goicovic2017,Goicovic2018}. Fortunately, such 
a configuration of accretion onto binary black holes has been found in \NGC\, by 
ALMA  \citep{Impellizzeri2019}, which provides unique opportunity to investigate the
orbital evolution predicted by simulations.  

{\it Signatures of CB-SMBHs.}
The suggested signatures so far are: 1) periodicity of long term radio or optical variations 
of AGNs from a few years to a few tens of years, which is regarded as results of modulation 
of the orbital motion of CB-SMBHs; 2) ultraviolet deficit of 
continuum rising from the central cavity of the CBD governed by the tidal torques of the 
binary; 3) shifts of red and blue peaks of broad emission lines in long term variations 
similar to normal binary stars; 4) 
deficit profile of brightness of galactic centers as a result of ejection of stars
through interaction of stars with the binary black holes; 5) 2-dimensional kinematic 
maps of broad emission lines in AGNs showing signatures of two point potentials generated
through reverberation mapping campaigns; 6) GRAVITY (or Extremely Large Telescope: ELT)
with spectroastrometry may reveal some signals of double BLRs. Additionally,
X-shaped radio jet was regarded as a signature of CB-SMBHs \citep{Merritt2002}, but it hardly 
applies to measure orbital parameters of the binaries.

Great efforts have been made to search for periodicity of AGN long-term variations over several 
decades. Nowadays, a sample composed of about 100 AGNs was built up through Catalina Real-time
Transient \citep{Graham2015a}, among which PG\,1302-102 is the best one with a period 
of 5\,yrs \citep{Graham2015b,DOrazio2015}. Other three objects with periodical variations are 
OJ\,287 in radio \citep{Valtonen2008}, NGC\,5548 with a period of about $\sim 13\,$yrs over
40\,yrs \citep{Li2016}, and Akn\,120 with 
a period of $\sim 20\,$yrs over the last 40\,yrs \citep{Li2019}. Progress in this way is quite
slow so far unless the future LSST can efficiently discover some AGNs with short periods
from the fast large scale of time domain surveys. 

Continuum features of CB-SMBHs radiated from accretion result from the cavity of the CBD
governed by tidal interaction \citep{Gultekin2012}. However, the shapes of UV continuum
are easily affected by dust extinctions \citep{Leighly2016}, making that it is hardly useful 
to justify candidates of CB-SMBHs in practice. Moreover, it has been 
shown that the binary black holes are peeling off gas from the inner edge of the CBD
making the accretion rates quite high (the rates still comparable with the accretion 
rates of the CBD) \citep{Farris2014,Shi2016,Bowen2019}. For the current case of \NGC,
the binary black holes are peeling off gas for accretion to radiate gravitational energies.

Searching for systematically periodical shifts of AGNs with double-peaked profiles is 
expected to find CB-SMBHs \citep{Runnoe2017,Doan2020}. 
Similar to classical binary stars, the double BLRs orbiting around the mass center
of the binary black holes 
will lead to opposite shifts of the red and blue peaks with orbital phases. The
long term campaigns are expected to carry out in next a few years, but
SDSS J0938+0057, SDSS J0950+5128, and SDSS J1619+5011 
in this sample showed systematic and monotonic velocity changes consistent with 
the binary hypothesis \citep{Runnoe2017}.

As a consequence of interaction with CB-SMBHs, as shown by numerical 
simulations \citep{Ebisuzaki1991,Milosavljevic2002,Merritt2006}, stars are ejected from galactic 
center, forming a ``mass deficit'' in its central part. Actually, it does not appear to 
have a pronounced and unambiguously detectable effects on the density profile. Even though
it works for searching for CB-SMBHs, on the other hand, this method doesn't apply to determine 
orbital parameters to test properties of gravitational waves, and the same shortcomings for the 
identifications of AGNs with X-shaped jet morphology.

Reverberation mapping of AGNs is a powerful tool to probe the kinematics and structure of 
the BLR \citep{Peterson1993}. Usually, BLRs have relatively stable structures justified from the 
large sample of SDSS quasars \citep{Berk2001} though they are variable on timescale of years. 
In principle, the broad-line gas is governed by the central black hole potential, and therefore
it is expected to show signatures of CB-SMBH orbital motion from the 2-dimensional velocity 
delay map from reverberation mapping of AGNs \citep{Wang2018,Songsheng2020,Kovacevic2020}. 
This reasoning also depends on the orbital motion, but the advantage of this scheme is 
obvious that the campaigns can be performed in seasons and does not require long-term monitoring 
as the approach of period searching. A campaign called as Monitoring AGNs with H$\beta$ Asymmetry 
(MAHA) has been conducted with Wyoming Infrared Observatory (WIRO) Telescope targeting on 
objects with various H$\beta$ profiles \citep{Du2018}. Results are expected to be carried out 
shortly.

GRAVITY/VLTI provides the highest spatial
resolution in NIR. The differential phase curves measured by GRAVITY depends on the
geometric structures and kinematics of ionized gas emitting broad lines. For CB-SMBHs,
there are several simplest configurations of AM distributions of individual BLRs and
binary orbital motion. Details of the phase curves have been explored for signals of 
CB-SMBHs \citep{Songsheng2019}. We are expecting to jointly observe some Seyfert 2 
galaxies through ALMA and GRAVITY, which can be selected  by 
checking optical polarized spectra with complicated profiles. We hope discover more
Seyfert 2 galaxies with the CRDs like in \NGC.

Some AGNs show that rotation curves of masers don't follow the Keplerian law. The radial
self-gravity of the maser  disc could be important \citep{Hure2011}, however, infrared emissions 
doesn't support this hypothesis. Moreover, ALMA measurements of molecular gas mass in \NGC\, 
is significantly smaller than the black hole. It is a curiosity that the maser  disc shares 
different dynamics from the molecular gas if inspecting 
details \citep{Greenhill1997,Impellizzeri2019}. 
On the other hand, it could be plausible, if speculate, that CB-SMBHs work in these objects. 
Detailed dynamical modeling of the maser or molecular gas at parsec scales will allow us to 
determine some of orbital parameters, which are keys to test properties of nano-Hz 
gravitational waves.

\section{Cavity of the CBD}
Dynamics of gas at parsec scales revealed by the water megamaser \citep{Greenhill1996,Gallimore1996a}
and HCN/HCO$^{+}$\,($J=3-2$) molecular gas \citep{Imanishi2018,Impellizzeri2019,Garcia-Burillo2019} 
provides strong constraints on the accretion  disc size of the central black hole and hence spatial 
distribution of gas in the compact regions. For \NGC, the black hole mass is of 
$\bhm=(0.8-1.7)\times 10^{7}\sunm$ and the bolometric luminosity is 
$L_{\rm Bol}\approx 2.0\times 10^{45}\,\ergs$ from multiwavelength data \citep{Gravity2020}, 
we thus have the Eddington ratio of $\lambda_{\rm Edd}\approx 1$,
implying that \NGC\, has high accretion rates. 
The strong \feii\, emissions in the polarized optical spectra \citep{Miller1991} shows 
the key feature of high rates \citep{Du2019}. 

Recent ALMA observations of HCN/HCO$^{+}$ ($J=3-2$) emission 
lines \citep{Impellizzeri2019,Garcia-Burillo2019} show that the extrapolation of the rotation 
curve of the $0.5-1.2\,$pc HCN molecular gas, which agrees with that of the water 
maser \citep{Greenhill1996,Gallimore1996b}, is consistent with the rotation curve extended to 
7\,pc  disc \citep{Impellizzeri2019}. This indicates 
that the enclosure mass is still dominated by the central black hole at a few 
parsec scales \citep{Hure2002}. With help of the gas  disc mass ($M_{\rm AD}$) estimated from 
Equation (\ref{eq: disc-mass}),  we have upper limit of its outer boundary of 
$R_{\bhm}\le 
1.95\times 10^{5}\,\left(\alpha_{0.1}M_{7}\right)^{16/25}\dotm_{10}^{-14/25}\,\Rs\approx0.19\,$pc
from $M_{\rm AD}(\le R)\le \bhm$. With the gas dynamics, we draw a conclusion that the  disc 
must be truncated at least $R_{\bhm}$. Indeed, the HCN molecular gas is 
$M_{\rm gas}^{\oplus}\sim 9\times 10^{5}\sunm$ ($\lesssim3\,$pc: corresponding to a surface 
density $\Sigma_{\rm gas}^{\oplus}\approx 3\times 10^{4}\,\sunm{\rm pc^{-2}}$ and a number 
density of $n_{\oplus}\approx 2.2\times10^{6}\,{\rm cm^{-3}}$) 
measured by ALMA \citep{Imanishi2018,Garcia-Burillo2019}, which is smaller than $\bhm$.
At this edge of the  disc, its temperature is $T_{\rm eff}(R_{\bhm})\approx 
120\,\left(\dotm_{10}/M_{7}\right)^{1/4}\left(R/R_{\bhm}\right)^{-3/4}\,$K
much lower than the dust sublimation temperature, and the  disc surface density is
$\Sigma_{\rm AD}\approx 5.7\times 10^{7}\,\sunm{\rm pc^{-2}}$. There is another
critical radius corresponding to the dust sublimation temperature, we have 
$R_{\rm dust}=0.04\,\left(\dotm_{10}/M_{7}\right)^{1/3}R_{\bhm}$.
According to the Kennicutt-Schmidt law \citep{Kennicutt1998}, this high surface density of 
gas between $(R_{\rm dust},R_{\bhm})$ with low temperatures are undergoing starbursts with a 
rate of $\calR\approx 4.8\,\sunm{\rm yr^{-1}}$, resulting in an infrared luminosity 
of $L_{\rm MIR}^{*}\approx 1.1\times 10^{44}\ergs$ radiating between 8-100$\mu{\rm m}$. Such 
a bright mid- and far infrared emission doesn't appear in the spectral energy distributions 
(the observed MIR luminosity is $L_{\rm MIR}^{\rm obs}\lesssim 1.0\times10^{43}\ergs$ estimated 
from Figure 5 in Ref\, \citep{Gravity2020}). Considering the constraints of 
$L_{\rm MIR}^{*}<L_{\rm MIR}^{\rm obs}$, we have the outer boundary radius
$R_{\rm AD}\lesssim\left(L_{\rm MIR}^{\rm obs}/L_{\rm MIR}^{*}\right)R_{\bhm}
\lesssim 0.1\,R_{\bhm}=0.019\,$pc, which is consistent with $R_{\rm dust}$. 
This means that the $R_{\rm AD}-R_{\bhm}$ region should be a cavity of gas (or a
low-density region though we are not able to estimate its density currently). On the 
other hand, GRAVITY \citep{Gravity2020} measured the radius of the dust sublimation 
$R_{\rm sub}\approx 0.24\pm0.03\,$pc and discovered an extended structure with a 
central hole with the radius of $R_{\rm sub}$ by imaging the central regions of 
\NGC\, at 2$\mu{\rm m}$, agreeing with $R_{\bhm}\lesssim R_{\rm sub}$. All these 
provide evidence for that the 2$\mu$m-imaged hole is a central cavity of gas.

Such a cavity is a natural consequence of tidal interaction of the circumnuclear  disc 
(CND) with the binary black 
holes  \citep{Artymowicz1994,Escala2005,MacFadyen2008,Cuadra2009,Lodato2009} if they 
exit subsequently evidenced by the CRD in the galactic center of \NGC. 
According to analytical analysis, the inner radius of the CND, which is denoted the 
circumbinary  disc (CBD), should be around 
$R_{\rm CBD}\approx R_{\rm sub}=2A$, where $A$ is the separations of the binary 
black holes. From the 2$\mu$m-imaged hole, we have $A\approx 0.12\,$pc.

\section{Accretion  discs}
Given the black hole mass of $\bhm=10^7\sunm$ in \NGC, distance scales are: 
$0.1\,{\rm pc}\approx 114\,{\rm ltd}\approx10^{5}\,\Rs$. The dimensionless accretion rate is defined 
as $\dotm=\dot{M}/L_{\rm Edd}c^{-2}= 10\,\eta_{0.1}^{-1}\lambda_{\rm Edd}$ from 
$L_{\rm Bol}=\eta\dot{M}c^{2}$, where $\eta_{0.1}=\eta/0.1$ is the radiative efficiency, 
$L_{\rm Edd}$ is the Eddington luminosity, $\lambda_{\rm Edd}=L_{\rm Bol}/L_{\rm Edd}$
is the Eddington ratio and $\dot{M}$ is the accretion rates.
In the outer part of the Shakura-Sunyaev  disc \citep{Shakura1973}, 
where gas pressure dominates over radiation pressure and absorption over Thompson scattering,
the surface density is given by
\begin{equation}
\Sigma_{\rm d}
    =1.88\times 10^{7}\,\alpha_{0.1}^{-4/5}M_{7}^{1/5}\dotm^{7/10}r_{5}^{-3/4}\,\sunm\,{\rm pc^{-2}},
\end{equation}
and the total mass of the accretion  disc within $R$,
\begin{equation}\label{eq: disc-mass}
M_{\rm AD}(\le R)\approx 8.67\times 10^{5}\,\alpha_{0.1}^{-4/5}M_{7}^{1/5}\dotm^{7/10}r_{5}^{5/4}\,\sunm.
\end{equation} 
In light of the Toomre's parameter, the accretion  disc becomes vertically self-gravitating at 
a radius of $R_{\rm SG}/\Rs\approx 500\,\alpha_{0.1}^{2/9}M_{7}^{-2/9}\dotm^{4/9}$ \citep{Laor1989}. 
However, we are interested in the constraints 
on the AD size due to radial self-gravitating effects in this paper. Viscosity dissipates gravitational 
energy leading to effective temperature of  disc surfaces
\begin{equation}
T_{\rm eff}=\left(\frac{3}{8\pi}\frac{G\bhm\dot{M}}{\sigma_{\rm SB}R^{3}}\right)^{1/4}
           =111.0\,M_{7}^{-1/4}\dotm^{1/4}r_{5}^{-3/4}\,{\rm K},
\end{equation}
where $\sigma_{\rm SB}$ are the Stefan-Boltzmann constant, respectively.

\end{document}